\documentclass [12pt]{article}
\usepackage{amssymb,amsmath,epsf,graphics}
\def\ep{\epsilon}
\def\lam{\lambda}

\def\dl{\delta}

\def\s2{\sigma^2}

\def\om{\omega}
\def\o2{\omega^2}
\def\al{\alpha}
\def\te{\theta}

\def\k2{\kappa^2}
\def\bfs{\boldsymbol}
\usepackage{epsfig}
\usepackage{amssymb}
\begin{document}
\title{ \bf{A skew Gaussian decomposable\\ graphical model}}
\author{\vspace {.2cm} {\normalsize Hamid  Zareifard$^a$, H{\aa}vard Rue$^b$,
Majid Jafari Khaledi$^c$, Finn Lindgren$^d$}\\ \\
$^a$ {\scriptsize    Department of Statistics, Jahrom University, Jahrom, Iran} \\
{\scriptsize zareifard@jahrom.ac.ir}\\
$^b$ {\scriptsize  Department of Mathematical Sciences, Norwegian
University of Science and Technology, N-7491 Trondheim, Norway} \\
{\scriptsize hrue@math.ntnu.no}\\
$^c$ {\scriptsize  Department of Statistics, Tarbiat Modares University, P.O. Box 14115-134, Tehran, Iran} \\
{\scriptsize  jafari-m@modares.ac.ir} \\
$^d$ {\scriptsize  Department of Mathematical Sciences, University of Bath, Claverton Down, Bath, BA2 7AY, United Kingdom} \\
{\scriptsize  f.lindgren@bath.ac.uk} }
\date{}
\maketitle
\begin{abstract}
This paper propose a novel decomposable graphical model to
accommodate skew Gaussian graphical models. We encode conditional
independence structure among the components of the multivariate
closed skew normal random vector by means of a decomposable graph
and so that the pattern of zero off-diagonal elements in the
precision matrix corresponds to the missing edges of the given
graph. 
We present conditions that guarantee the propriety of the posterior
distributions under the standard noninformative priors for mean
vector and precision matrix, and a proper prior for skewness
parameter. The  identifiability of the parameters is investigated by
a simulation study. Finally, we apply our methodology to two data
sets.

\end{abstract}
\section{Introduction}
In recent years, there have been many developments in multivariate
statistical models. Making sense of all the many complex
relationships and multivariate dependencies present in the data,
formulating correct models and developing inferential procedures is
an important challenge in modern statistics. In this context,
graphical models currently represent an active area of statistical
research which have served as tools to discover structure in  data.
More specifically, graphical models are multivariate statistical
models in which the corresponding joint distribution of a family of
random variables is restricted by a set of conditional independence
assumptions, and the conditional relationships between random
variables are encoded  by means of a graph. In the Gaussian case,
these models  induce the conditional independence assumptions by
zeros in the precision matrix. An important reason for working with
this class of distributions is important properties like closure
under marginalization, conditioning and linear combinations which is
seldom preserved outside the class of multivariate normal
distributions. However, in spite of substantial advances,  the
Gaussian distributional assumption might be overly restrictive to
represent the data. The real data could be highly non-Gaussian and
may show features like skewness.

In this article, we study skew distributions in graphical models
with the aim of mimicking the success of Gaussian graphical
models as much as possible. 
The last decade has witnessed major
developments in models whose finite dimensional marginal
distributions are multivariate skew-normal. Azzalini and Capitanio
(1999) introduced multivariate skew-normal (SN) distribution which
enjoys some of the useful properties of normal distribution, such as
property of closure under marginalization and
conditioning.  
Accordingly, an $n$-dimensional random vector ${\bf{Y}}$ is said to
have a SN distribution if its density is
\begin{eqnarray*}
\phi_n({\bf{y}};{\bfs\mu},\Omega)\Phi(\al_0+{\bf\al}'D_\Omega^{-1}({\bf{y}}-{\bfs\mu}))/
\Phi(\tau),
\end{eqnarray*}
where $\phi_n(\cdot;{\boldsymbol{\mu}},\Omega)$ is the probability
density function of the $n$-dimensional
$N_n({\boldsymbol{\mu}},\Omega)$ variable, $\Phi(\cdot)$ is
cumulative distribution function  of $N(0,1)$,
${\boldsymbol{\mu}}\in\Re^n$, $\tau\in\Re$, $\Omega\in\Re^{n\times
n}$ is a full rank covariance matrix,
$D_\Omega=diag(\Omega_{11},\cdots,\Omega_{nn})^{1/2}$,
${\bfs\al}\in\Re^{n}$ is shape parameter and
$\al_0=\tau(1+{\bfs\al}'D_\Omega^{-1}\Omega
D_\Omega^{-1}{\bfs\al})^{1/2}$.   When ${\bfs\al}={\bf0}$ we are
back  to  the  multivariate normal distribution. Capitanio et al.
(2003) used the SN family in graphical models examining in
particular the construction of conditional independence graphs.
Their results show that if ${\bf Y}$ be an n-variate SN distribution
with covariance matrix $\Omega$ and skewness vector ${\bfs\al}$,
then
\begin{eqnarray*}
Y_i\perp Y_j| {\bf Y}_{-ij}\ \ \Leftrightarrow\ \ \Omega^{ij}=0 \ \
and\ \ \al_i\al_j=0
\end{eqnarray*}
where  ${\bf Y}_{-ij}$ is ${\bf Y}$ with the $i$th and $j$th
elements deleted and $\Omega^{ij}$ denotes the $(i,j)$th entry of
$\Omega^{-1}$. Comparing with the Gaussian graphical model, an extra
constraint $\al_i\al_j=0$ is necessary to capture conditional
independence property. It means if we believe $Y_i$ and $Y_j$ are
conditionally independent then at least one of $\al_i$ and $\al_j$
must be zero. Hence, applying this constraint in practical issues is
challenging. Alternatively, Dominguez-Molina et al. (2003) and
Gonzalez-Farias et al. (2004) proposed the multivariate closed skew
normal (CSN) distribution which  includes the property of SN family.
Also unlike the SN family, the CSN family enjoys this property that
the joint distribution of i.i.d. CSN random variables is the
multivariate CSN distribution.

Although much progress has been made in the context of skew normal
distributions,  the achieved successes  in graphical models are
limited. The preservation of conditional independence property for
skew normal variables has shown that the extending the class of
skew-normal distributions to graphical models is challenging. The
aim of this paper is to develop a multivariate closed skew normal
graphical model. We encode conditional independence structure among
the components of the multivariate closed skew normal random vector
with respect to a decomposable graph $G$. The main motivation to use
the decomposable graphs for encoding the conditional independence is
that for this type of graph there exists an ordering of the vertices
such that the zero elements in precision matrix are reflected in its
Cholesky decomposition (Paulsen et al., 1989). Under decomposable
graphs, the conditional independence property is maintained for our
skewed graphical model, and simplification occurs in both the
interpretation of data and the estimation procedure. Models can be
specified in terms of conditional and marginal probability
distributions, leading to a simplified analysis based on lower
dimensional components (Giudici and Green, 1999; Letac and Massam,
2007; Khare and Rajaratnam, 2011).

The precision matrix is the fundamental object that evaluates
conditional dependence  between random variables. Estimating a
sparse precision matrix  is crucial specially in high-dimensional
problems. In this context, a family of conjugate prior distributions
for the precision matrix is developed. Conditions for propriety of
the posterior  are given under the standard noninformative priors on
mean vector and precision matrix as well as a proper prior for the
skewness parameter. We also develop and implement a Markov chain
Monte Carlo (MCMC) sampling approach for inference.

The organization of the paper is as follows. Section 2 introduces
the required preliminaries and notation. In Section 3, a novel skew
Gaussian decomposable graphical model is  constructed using a
multivariate closed skew normal distribution and  its properties are
established.  Section 4 discusses Bayesian analysis using Gibbs
sampling to sample from the posterior distribution.  A simulation
study is reported in Section 5. Section 6 illustrates the use of
proposed methodology in two real data sets: an analysis of student
marks from Mardia et al. (1979) and  an analysis of the carcass data
from gRbase package of R. Finally, conclusions and discussion are
given in Section 7. The Appendix  contains proofs of some of the
results  in the main text.
\section{Preliminaries}
\subsection{Multivariate closed skew-normal distribution}
An $n$-dimensional random vector ${\bf{Y}}$ is said to have a
multivariate closed skew-normal distribution, denoted by
$CSN_{n,m}({\boldsymbol{\mu}},\Sigma,\Gamma,{\boldsymbol{\nu}},\Delta)$,
if its density function is of the form
\begin{eqnarray}
f({\bf y})=\phi_n({\bf{y}};{\boldsymbol{\mu}},\Sigma)
\Phi_m(\Gamma({\bf{y}}-{\boldsymbol{\mu}});{\boldsymbol{\nu}},\Delta)
/\Phi_m(0;{\boldsymbol{\nu}},\Delta+\Gamma\Sigma\Gamma'),\label{1}
\end{eqnarray}
where ${\boldsymbol{\mu}}\in\Re^n$, ${\boldsymbol{\nu}}\in\Re^m$,
and $\Sigma\in\Re^{n\times n}$ and $\Delta\in\Re^{m\times m}$ are
both covariance matrices, $\Gamma\in\Re^{m\times n}$, and
$\Phi_n(\cdot;{\boldsymbol{\mu}},\Sigma)$ is the cumulative
distribution function of the $n$-dimensional normal distribution
with mean vector ${\boldsymbol{\mu}}$ and covariance matrix
$\Sigma$.  To derive this distribution, Gonzalez-Farias et al.
(2004) consider a $(n+m)$-dimensional normal random vector
\begin{eqnarray*}
(W_{0_1},\cdots,W_{0_m},W_1,\cdots,W_n)'=\left(%
\begin{array}{c}
  {\bf W_0} \\
  {\bf W} \\
\end{array}%
\right)\sim N_{n+m}\left(%
\begin{array}{cc}
  {\bf 0},&\left(%
\begin{array}{cc}
  \Delta+\Gamma\Sigma\Gamma' & \Gamma\Sigma \\
  \Sigma\Gamma' & \Sigma \\
\end{array}%
\right) \\
\end{array}%
\right).
\end{eqnarray*}
Then, the probability density function of ${\bf Y}=({\bf W}|{\bf
W}_0>{\bfs\nu})$  is the multivariate closed skew-normal (\ref{1}).
If $\Gamma=0$ and $m=1$, this density reduces to the multivariate
normal one and the skew-normal distribution (Azzalini, 2005),
respectively. Allard and Naveau (2007) used the multivariate closed
skew normal and introduced a spatial skewed Gaussian process as a
novel way of modeling skewness for spatial data. To increase the
amount of skewness in the vector ${\bf{Y}}$ as well as to simplify
the interpretation of this density, they assumed that $m=n$, ${\bfs
\nu=0}$, $\Delta=\Sigma$ and $\Gamma=\alpha I_n$ in which $\alpha\in
\Re$ is a single parameter controlling skewness and $I_n$ is the
identity matrix of order $n$. This model is also referred to as the
homotopic model.

We will make use of the alternative representation of
Dominguez-Molina et al. (2003) for CSN families,
\begin{eqnarray}
\label{doming}{\bf{Y}}\stackrel{d}{=}{\bfs\mu}+(\Sigma^{-1}+\Gamma'\Delta^{-1}\Gamma)^
{-\frac{1}{2}}{\bf V}+
\Sigma\Gamma'(\Delta+\Gamma\Sigma\Gamma')^{-1}{\bf U},
\end{eqnarray}
where  ${\bf V}$ and $\bf{U}$ are independent multivariate normal
distribution $N_n(0,I_n)$ and a multivariate truncated normal
distribution $TN_n(0;0,\Delta+\Gamma\Sigma\Gamma')$, respectively,
where $TN_n({\bf{c}};{\boldsymbol{\mu}},\Sigma)$ denotes the
$N_n({\boldsymbol{\mu}},\Sigma)$ distribution truncated below at the
vector ${\bf{c}}$. Also, if ${\bf Y}$ is partitioned as ${\bf Y}=(
{\bf Y}'_1, {\bf Y}_2')'$, then the conditional distribution of
$(n-k)$-dimensional vector ${\bf Y}_2$ given ${\bf Y}_1={\bf
y}_{1_0}$ is
\begin{eqnarray*}
CSN_{n-k,m}({\bfs\mu}_2+\Sigma_{21}\Sigma_{11}^{-1}({\bf
y}_{1_0}-{\bfs\mu}_1)
,\Sigma_{22.1},\Gamma_2,{\bfs\nu}-\Gamma^*({\bf
y}_{1_0}-{\bfs\mu}_1),\Delta),
\end{eqnarray*}
where
$\Sigma_{22.1}=\Sigma_{22}-\Sigma_{21}\Sigma_{11}^{-1}\Sigma_{12}$,
$\Gamma^*=\Gamma_1+\Gamma_2\Sigma_{21}\Sigma_{11}^{-1}$ and the
parameters are induced corresponding to the partition of ${\bf Y}$
as
\begin{eqnarray*}
{\bfs\mu}=\left(%
\begin{array}{c}
  {\bfs\mu}_1 \\
  {\bfs\mu}_2 \\
\end{array}%
\right), \ \ \ \ \ \ \Sigma=\left(%
\begin{array}{cc}
  \Sigma_{11} & \Sigma_{12} \\
  \Sigma_{21} & \Sigma_{22} \\
\end{array}%
\right), \ \ \ \ \ \ \Gamma=\left(%
\begin{array}{cc}
  \Gamma_1 & \Gamma_2 \\
\end{array}%
\right).
\end{eqnarray*}

\subsection{Graph theory}
Let $V$ be a finite set of vertices and $E=\{(u,v): u,v\in V, u\neq
v\}$ be a set of edges so that $E\subseteq V\times V $. Define a
graph $G$ as an ordered pair $G=(V,E)$ of vertices and edges, where
$V$ is assumed to be finite. When $(u,v)\in E$, we say that $u$ and
$v$ are adjacent in $G$. A graph is said to be complete if all the
vertices are adjacent to each other. It is  understood that
$(u,v)\in E$  implies $(v,u)\in E$, i.e., the edges are undirected.
For an undirected graph $G=(V,E)$,  we say that $u$ and $v$ are
neighbors when $(u,v)\in E$. For any $A\subset V$, a subgraph of $V$
is defined as the graph $G_A=(A,E\cap (A\times A))$. A path from
$v_1$ to $v_m$ is a sequence of distinct vertices in $V$,
$v_1,v_2,\cdots,v_m$, for which $(v_j, v_{j+1})\in E$ for
$j=1,\cdots,m-1$. A subset $C\subset V$ separates two vertices
$i\notin C$ and $j\notin C$, if every path from $i$ to $j$ contains
at least one vertex from $C$. A clique of $G$ is a  complete
subgraph of $G$. A subgraph is  a maximal clique if it is not
contained in a larger complete subgraph. Any path that begins and
ends at the same vertex is called a cycle. A tree is a connected
graph with no cycles.\\
\\
{\bf Definition 1: (Decomposable graph (Lauritzen, 1996))} An
undirected graph is said to be decomposable if any induced subgraph
does not contain a cycle of length greater than or equal to four.

\begin{figure}
\centerline{
\includegraphics[width=130mm,height=43mm]{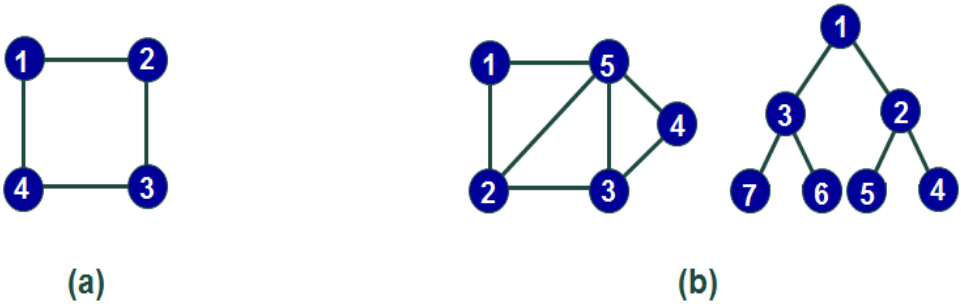}}
\caption{(a) A non-decomposable graph, and (b) two decomposable
graphs.}\label{decompose}
\end{figure}

Figure  \ref{decompose} shows  examples of non-decomposable and
decomposable graphs. Decomposable graphs have
several characterizations in terms of vertex orderings.\\
\\
{\bf Definition 2:} For an undirected graph $G=(V,E)$, an ordering
$\gamma:V\mapsto\{v_1,\cdots,v_m\}$ of $m$ vertices $(1,\cdots,m)$
is known as a perfect vertex elimination scheme for G if for every
triplet $v_i,v_j,v_k$ with $1\leq i<j<k\leq m$ the following
condition holds:
\begin{eqnarray*}
(\gamma^{-1}(v_j),\gamma^{-1}(v_i))\in E,\ \
(\gamma^{-1}(v_k),\gamma^{-1}(v_i))\in E\ \ \Rightarrow\ \
(\gamma^{-1}(v_k),\gamma^{-1}(v_j))\in E.
\end{eqnarray*}
In Figure \ref{decompose}(b), we show examples of perfect vertex
elimination schemes for some decomposable graphs. The existence of
such an ordering is an important advantage of decomposable over
nondecomposable graphs, and  the existence of this ordering
characterize decomposable graphs. More formally, every decomposable
graph admits an ordering of vertexes in terms of its cliques which
is a perfect vertex elimination scheme. If for an undirected graph
$G$, there exists an ordering  of its vertices corresponding to a
perfect vertex elimination scheme, then $G$ is a decomposable graph
(Lauritzen, 1996, page 18). However, this ordering need not be
unique. A constructive way to obtain such an ordering is given in
Lauritzen (1996).
\\
\\
{\bf Definition 3: (Modified Cholesky decomposition)} If $Q$ is a
positive definite matrix, then there exists a unique decomposition
\begin{eqnarray*}
Q=L'DL
\end{eqnarray*}
where $L$ is a upper triangular matrix with unit diagonal entries
and $D$ a diagonal matrix with positive diagonal entries. The
Cholesky decomposition of $Q$ is $\Gamma'\Gamma$   where
$\Gamma=D^{\frac{1}{2}}L$ is called the  Cholesky triangle.

The following lemma  will play a central role in our work.\\
\\
{\bf Lemma 1: (Paulsen et al. 1989)} Let $\Omega$ be an arbitrary
positive definite matrix with zero restrictions according to
decomposable graph $G=(V,E)$, i.e., $\Omega_{ij}=0$ whenever
$(i,j)\notin E$. Then there exists an ordering of the vertices such
that if $\Omega= L'DL$ is the modified Cholesky decomposition
corresponding to this ordering, then for $i<j$,
\begin{eqnarray*}
L_{ij}=0\Leftrightarrow (i,j)\notin E.
\end{eqnarray*}
Hence, the zeros in $\Omega$ are preserved in the lower triangle of
the corresponding matrix $L$ obtained from the modified Cholesky
decomposition. We  assume from now on that graph $G$ is
decomposable.

An $m$-dimensional Gaussian  graphical model can be represented by
the class of multivariate normal distributions with fixed zeros in
the precision matrix  (i.e., conditional independencies) described
by a given graph $G = (V,E)$ where the number of vertices is $m$.
That is, if $(i,j)\notin E$, the $i$th and $j$th components of the
multivariate random vector are conditionally independent.

We now introduce some notations and spaces from Khare and Rajaratnam
(2011). Let $M_m^+$ and $M_m$ denote the cone of symmetric positive
definite matrices of order $m$ and symmetric positive semidefinite
matrices of order $m$, respectively, then  the two parameter sets
$P_G^m$ and $SP_G^m$ according to decomposable graph $G=(V,E)$ are
defined as
\begin{eqnarray*}
P_G^m&=&\{\Omega\in M_m^+\ |\ \Omega_{ij}=0, \ \ (i,j)\notin E\},\\
SP_G^m&=&\{\Omega\in M_m\ |\ \Omega_{ij}=0, \ \ (i,j)\notin E\},
\end{eqnarray*}
and also define
\begin{eqnarray*}
\ell_G^m&=&\{L:\ \ L_{ij}=0 \ \ whenever\ i>j,\ or \ (i,j)\notin E\
and\
L_{ii}=1 \ \forall \ 1\leq i,j\leq m \},\\
\Theta_G^m&=&\{\te=(L,D):\ \ L\in\ell_G^m \ \ and\ D\ diagonal\
with\ D_{ii}>0\  \forall\  1\leq i \leq m \}.
\end{eqnarray*}
\section{Skew Gaussian decomposable graphical models}
To introduce a skew version of $k$-dimensional Gaussian  graphical
(GG) model,  we first consider the following tree graphical model in
a loose and imprecise form
\begin{eqnarray}
{\stackrel{1}{\bigcirc}}\line(1,0){30}\stackrel{2}{\bigcirc}\line(1,0){30}\stackrel{3}{\bigcirc}.\label{lin.g}
\end{eqnarray}
Based on this graph, let ${\bf X}=(X_1,X_2,X_3)$ be a random vector
whose elements are indexed by graph (\ref{lin.g}). We introduce
three independent increments
\begin{eqnarray*}
\ep_1=\kappa_1X_1,\ \ \ \ \ep_2=\kappa_2(X_2-b_{12}X_1),\ \ \ \
\ep_3=\kappa_3(X_3-b_{23}X_2).
\end{eqnarray*}
where $\kappa_1,\kappa_2,\kappa_3\in\Re^+$ and
$b_{12},b_{23}\in\Re$. Now, we can relate
${\bfs\ep}=(\ep_1,\ep_2,\ep_3)$ to ${\bf X}=(X_1,X_2,X_3)$ by
${\bfs\ep}=D_\kappa^{\frac{1}{2}} B{\bf X}$ where
\begin{eqnarray*}
B=\left(%
\begin{array}{ccc}
  1 & 0 & 0 \\
  -b_{12} & 1 & 0 \\
  0 & -b_{23} & 1 \\
\end{array}%
\right),\ \ \ \ \ and \ \ \ \
D_\kappa=diag(\kappa_1^2,\kappa_2^2,\kappa_3^2).
\end{eqnarray*}
If we suppose $\ep_1,\ep_2,\ep_3\stackrel{iid}{\sim} N(0,1)$, then
the joint density of ${\bf X}=(X_1,X_2,X_3)$ becomes  trivariate
normal distribution with fixed zeros in the precision matrix
described by given graph. Our aim  is to assess the suitability and
wider applicability of asymmetric distributions in graphical models
with the hope of mimicking the success of Gaussian graphical models
as much as possible. To do this, we need to study the following two
questions
\begin{itemize}
    \item How  can we choose a skewed distribution for $\bfs\ep$ such that the distribution of
    ${\bf X}$ belong to the same class as that of the $\bfs\ep$?
    \item What conditions are necessary to achieve the Markov property for the skewed graphical model?
\end{itemize}
To address these questions, we will use the CSN model that is more
general than SN, and is closed under marginalization, conditioning
and linear combination. Also unlike the SN family the joint
distribution of i.i.d. CSN random variables is a multivariate CSN
distribution. We now assume
\begin{eqnarray*}
\ep_i\sim CSN_{1,1}(0,1,\al_i,0,1),\ \ \ \ i=1,2,3\ \
\Leftrightarrow\ \  {\bfs\ep}\sim CSN_{3,3}({\bf 0},I_3,D_\al,{\bf
0},I_3)
\end{eqnarray*}
where $D_\al=diag(\al_1,\al_2,\al_3)$. Note that $\al_i$ is
interpreted as skewness parameters of the $i$th increment. Using the
closure property of the CSN distribution under linear
transformations, the joint density for ${\bf X}$ becomes
\begin{eqnarray*}
p({\bf x})=2^3\phi_3({\bf{x}};0,(B'D_\kappa B)^{-1})\Phi_3(D_\al
B{\bf{x}};{\bf 0},D_\kappa^{-1}).
\end{eqnarray*}
In this  example, we can easily show that $X_1\perp X_3|X_2$, so we
have the Markov property. The question is then: how can we extend
this easy example to a more general graphical model? This we will study next.\\
\\
{\bf Definition 4:} Consider an undirected  decomposable graph
$G=(V,E)$ where the number of vertices is $k$ and ordering of the
vertices corresponds to a perfect vertex elimination scheme. A
random vector ${\bf X}=(X_1,\cdots,X_k)'$ is called a skew Gaussian
decomposable graphical  (SGDG) model with respect to  graph $G$ with
mean ${\bfs\mu}\in\Re^k$, the precision matrix $Q\in P_G^k$ and the
skewness parameters $\bfs\al\in\Re^k$, if its density is
$CSN_{k,k}({\bfs\mu},Q^{-1},D_\al L,{\bf 0},D_\kappa^{-1})$  or
equivalently
\begin{eqnarray}
p({\bf x})&=&2^k\phi_k({\bf{x}};{\bfs\mu},Q^{-1})\Phi_k(D_\al
L({\bf{x}}-{\bfs\mu});{\bf
0},D_\kappa^{-1})\nonumber\\
&=& (\frac{2}{\pi})^{k/2}|D_\kappa|^{1/2} \exp({-\frac{1}{2} ({\bf
x}-{\bfs\mu})'Q ({\bf x}-{\bfs\mu})})\Phi_k(D_\al L({\bf
x}-{\bfs\mu});{\bf 0},D_\kappa^{-1}),\label{sgmrf.decom}
\end{eqnarray}
where $D_\kappa=diag(\kappa_1^2,\cdots,\kappa_k^2)$,
$D_\al=diag(\al_1,\cdots,\al_k)$ and $(L,D_\kappa)\in\Theta_G^k$ and
correspond to modified Cholesky decomposition of $Q$.
\begin{figure}
\centerline{
\includegraphics[width=167mm,height=85mm]{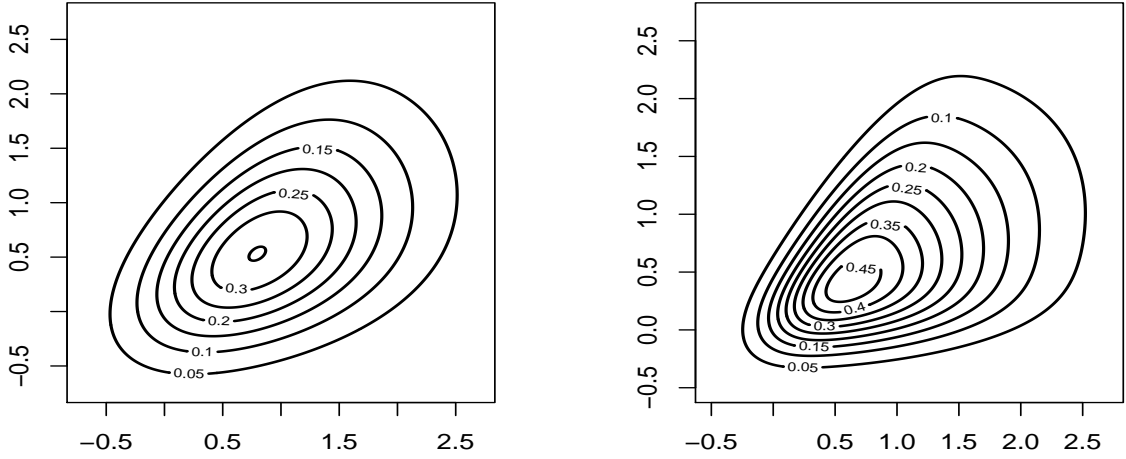}}
\caption{Left: contour plot of the SGDG model  with parameters
${\bfs\mu=(0,0)'}$, $D_\kappa=diag(1,1)$, $L_{12}=-0.5$ and
$\al_1=\al_2=2$. Right: contour plot of the SGDG model with
parameters ${\bfs\mu=(0,0)'}$, $D_\kappa=diag(1,1)$, $L_{12}=-0.5$
and $\al_1=\al_2=4$.}\label{contour}
\end{figure}

When skewness parameters are all zero, the density
(\ref{sgmrf.decom}) reduces to the multivariate normal. Figure
\ref{contour} shows contour plots of the SGDG model ($k=2$) for
different skewness parameters. Although Definition  4 depends on the
ordering of the vertices, this is not as restrictive as it first
appears. The ordering is essentially another parameter to be
specified and  can  be viewed as imposing extra information. In
sequel, we show that the SGDG model is supported
by the property of conditional independence.\\
\\
{\bf Theorem 1:} Let ${\bf X}$ be closed skew-normal distributed
corresponding to Definition 4. Then for $i< j$, we have
\begin{eqnarray*}
X_i\perp X_j|{\bf x}_{-ij}\ \ \Leftrightarrow\ \ Q_{ij}=0\ \
\Leftrightarrow\ \ L_{ij}=0\ \ \Leftrightarrow\ \ (i,j)\notin E.
\end{eqnarray*}

This is one of the main results. It simply says that similar to the
Gaussian graphical models the nonzero pattern of $Q$ determines $G$,
so we can read off from $Q$ whether $X_i$ and $X_j$ are
conditionally independent, so the natural way to parametrize the
SGDG model is by its precision matrix $Q$.  Although, the results in
this article are based on decomposable graphical models, in sequel
we have a theorem based on an arbitrary undirected graph. \\
\\
{\bf Lemma 2:} Consider an arbitrary undirected  graph $G=(V,E)$
where $V=\{1,2,\cdots,k\}$. Let $L$ be the modified cholesky
decomposition of the precision matrix corresponding to graph $G$.
Define the set $F(i,j)=\{i+1,\cdots,j-1,j+1,\cdots,k\}$. Now, if
$F(i,j)$ separates $i<j$ in $G$, then $L_{ij}=0$ (see Rue
and Held, 2005).\\
\\
{\bf Theorem 2:} Consider an arbitrary undirected graph $G=(V,E)$.
Let $L$ be the modified Cholesky triangle of precision matrix $Q\in
P_G^k$. Suppose ${\bf X}$ is closed skew-normal distributed
(\ref{sgmrf.decom}), then for $i<j$, we have
\begin{eqnarray*}
if\  F(i,j)\ separates\ two\ vertices\ i\ and\ j\ \ \Rightarrow \ \
X_i\perp X_j|{\bf x}_{-ij}.
\end{eqnarray*}

Alternatively, using the representation of Dominguez-Molina et al.
(2003) for CSN families,  a stochastic representation of the SGDG
model is
\begin{eqnarray}
{\bf
X}\stackrel{d}{=}{\bfs\mu}+L^{-1}D_\kappa^{-\frac{1}{2}}D_\al(I+D^2_\al)^{-\frac{1}{2}}{\bf
U} +L^{-1}D_\kappa^{-\frac{1}{2}}(I+D^2_\al)^{-\frac{1}{2}}{\bf
V},\label{linear}
\end{eqnarray}
where  ${\bf V}$ and ${\bf U}$ are independent multivariate normal
distribution $N_k({\bf 0},I_k)$ and a multivariate  half-normal
distribution $HN_k({\bf 0},I_k)$, respectively.  From a
computational point of view, this representation is useful because
it implies that an random vector distributed  according to
(\ref{sgmrf.decom}) can be generated using two independent  normal
random vectors. The mean vector and covariance matrix of ${\bf X}$
is
\begin{eqnarray*}
E({\bf X})&=&{\bfs\mu}+L^{-1}D_\kappa^{-\frac{1}{2}}D,\\
Cov({\bf
X})&=&(L'D_\kappa^{\frac{1}{2}}(I_k-D^2)D_\kappa^{\frac{1}{2}}L)^{-1}
\end{eqnarray*}
where diagonal matrix $D$ is
$D=\sqrt{\frac{2}{\pi}}D_\al(I_k+D^2_\al)^{-\frac{1}{2}}$. Hence,
the zero elements of inverse of  covariance matrix ${\bf X}$ are the
same of those of the precision matrix $Q$.
\section{Bayesian analysis}
In this section, we will discuss Bayesian inference of the SGDG
model (\ref{sgmrf.decom}). Assume $n$ independent observations ${\bf
x}=({\bf x}_1,\cdots,{\bf x}_n)$ from this model. We use the
representation (\ref{linear}) and  the following hierarchical
representation
\begin{eqnarray}
{\bf X}_j|{\bf U}_j&\sim&N_k({\bfs\mu}+L^{-1}D_\delta{\bf
U}_j,(L'D_\omega L)^{-1}), \ \ \ \ {\bfs\delta},\
{\bfs\mu}\in\Re^k,\ \ (L,D_\omega)\in\Theta_G
\nonumber\\
{\bf U}_j&\sim& HN_k({\bf 0},I_k),\ \ \ \ \ \
j=1,\cdots,n,\label{hirar}
\end{eqnarray}
where $D_\omega=diag(\omega_1^2,\cdots,\omega_k^2)$ and
$D_\delta=diag(\delta_1,\cdots,\delta_k)$ with
$\delta_i=\frac{\al_i}{\kappa_i\sqrt{1+\al_i^2}}$ and
$\omega_i={\kappa_i\sqrt{1+\al_i^2}}$. The later reparametrization
has been imposed to ease the computations. We will now discuss the
prior distribution for the unknown parameters
${\bfs\eta}=({\bfs\mu},{\bfs\delta},L,D_\omega)$. With regard to
relation between $\delta_i$ and $\om_i$ in the mentioned
reparametrization (i.e. $\delta_i=\frac{\al_i}{\om_i}$), we will use
a correlated  prior  for ${\bfs\delta}=(\delta_1,\cdots,\delta_k)$
and ${\bfs\om}=(\om_1,\cdots,\om_k)$;
\begin{eqnarray*}
\pi({\bfs\delta}|{\bfs\om})= N_k({\bf 0},b_1D_\om^{-1}),
\end{eqnarray*}
where  $b_1\in\Re_+$ is fixed. For the priors on ${\bfs\mu}\in\Re^k$
and $(L,D_\omega)\in\Theta_G$. We will discuss two separate prior
structures  corresponding to independent proper and noninformative
priors for these parameters.

{\bf Independent proper priors on ${\bfs\mu}$, ${\bfs\om}^2$ and
$L$}: We   take normal priors for mean vector and  lower triangular
matrix  and gamma prior for precision parameters as follows
\begin{eqnarray}
\pi({\bfs\mu})&=& N_k({\bfs\mu}_0,b_2I_k),\nonumber\\
\pi({\bfs\om}^2)&=& \prod_{i=1}^kG(b_3,b_4),\nonumber\\
\pi(L_{1.}^{\neq 0},\cdots,L_{k-1.}^{\neq
0})&=&\prod_{i=1}^{k-1}N_{||N^\prec(i)||}({\bfs
0},b_5I_{||N^\prec(i)||}), \label{indep}
\end{eqnarray}
where  $G(a,b)$ denotes the gamma distribution, $L_{i.}^{\neq
0}=(L_{ij})_{\{i<j,\ (i,j)\in E\}}$ denotes the nonzero off-diagonal
elements of $i$'th row of $L$ and $N^\prec(i):=\{j:\ (i,j)\in E,
i<j\}$. Also, ${\bfs\mu}_0\in\Re^k$ and $b_i\in\Re_+, i=2,3,4,5,$
are known hyperparameters.

{\bf Non-informative priors on ${\bfs\mu}$, ${\bfs\om}^2$ and $L$}:
Assign a common noninformative prior distributions for the mean
vector of the form $\pi({\bfs\mu})\propto { 1}$. Khare and
Rajaratnam (2011) formed a rich and flexible class of Wishart priors
for decomposable covariance graphs  in Gaussian covariance graph
models in which marginal independence among the components of a
multivariate random vector is encoded by means of a graph $G$. They
also considered the case where G is decomposable. Although
covariance graph models are distinctly different from the
traditional concentration graph models, after some modifications,
their prior is extendable in our problem case. In this context,  the
class of measures on $(L,D_\omega)\in\Theta_G$ is provided with
density
\begin{eqnarray}
\pi(L\in dL,D_\omega\in dD_\omega)\propto \prod_{i=1}^k
(\omega^2_i)^{\psi_i/2-1}e^{-\frac{1}{2}tr((L'D_\omega L)\Psi)}
\prod_{i<j,(i,j)\in E}dL_{ij}\prod_{i=1}^kd\omega^2_i,\label{khar}
\end{eqnarray}
where positive definite matrix $\Psi$  and $\psi_i\in\Re_+, \
i=1,\cdots,k$ are known hyperparameters.  Theorem 3 provides a
sufficient condition for the
existence of a normalizing constant for $\pi(L,D_\omega)$.\\
\\
{\bf Theorem 3:} Let $dL:=\prod_{i<j,(i,j)\in E}dL_{ij}$ and
$dD_\omega:=\prod_{i=1}^kd\omega^2_i$. if $\psi_i>||N^\prec(i)||$,
then
\begin{eqnarray}
\int_{\Theta_G}\prod_{i=1}^k
(\omega^2_i)^{\psi_i/2-1}\exp\{-\frac{1}{2}tr((L'D_\omega L)\Psi)\}
dD_\om dL<\infty.\label{noninfo}
\end{eqnarray}

A standard noninformative prior for lower triangular matrix $L$ and
precision parameters ${\bfs\omega}^2$ can be chosen by respecting
the zeros  for $\Psi$  and $\psi_i, \ i=1,\cdots,k$. 
Hence, the resulting non-informative prior on ${\bfs\mu}$,  $L$ and
${\bfs\om}^2$ is of the form
\begin{eqnarray}
\pi({\bfs\mu},L,{\bfs\om}^2)\propto\prod_{i=1}^k\frac{1}{\o2_i}.\label{gef0}
\end{eqnarray}
Theorem 4 shows that the posterior distribution is proper under
prior distribution (\ref{gef0}). To show this, we use a result by
Mouchart (1976) and Florens {\it{et al}}. (1990) which implies that
the posterior distribution exists as a proper only when
\begin{eqnarray*}
p({\bf x})=\int p({\bf
x}|{\bfs\mu},{\bfs\omega}^2,{\bfs\delta},L)\pi({\bfs\mu},{\bfs\omega}^2,{\bfs\delta},L)
d{\bfs\mu}d{\bfs\omega}^2d{\bfs\delta}dL<\infty.
\end{eqnarray*}
\\
{\bf Theorem 4:}  Under the standard noninformative prior in
(\ref{gef0}) and with $n$ independent replication from the
hierarchical model (\ref{hirar}), the posterior distribution of
parameters exists if $n\geq \max\{||N^\prec(i)||\}+2$.

We will now discuss how to generate samples from the posterior
distribution. To facilitate the sampling, we introduce the latent
variables ${{\bf U}}=({\bf U}_1,\cdots,{\bf U}_n)$, and then use
Gibbs sampling to generate samples. The block full conditionals are
as follows:
\begin{eqnarray*}
\pi({\bf{u}}|{\bf{x}},{\bfs\mu},{\bfs\delta},{\bfs\omega}^2,L)&=&\prod_{i=1}^n\prod_{j=1}^k
HN(\frac{\omega^2_j\delta_j}{1+\omega^2_j\delta_j^2}L_{j.}({\bf
x}_i-{\bfs\mu}),
\frac{1}{1+\omega^2_j\delta_j^2}),\\
\pi({\bfs\delta}|{\bf{x}},{\bf{u}},{\bfs\mu},{\bfs\omega}^2,L)&=&
N_k(\Sigma_\delta^{-1}[\Sigma_{i=1}^n{D_{u_i}L({\bf x}_i-{\bfs\mu})}],\Sigma_\delta^{-1}D_\omega^{-1}),\\
\pi({\bfs\mu}|{\bf{x}},{\bf{u}},{\bfs\delta},{\bfs\omega}^2,L)&=&
N_k(\Sigma_\mu^{-1}[(L'D_\omega L)\Sigma_{i=1}^n({\bf
x}_i-L^{-1}D_\delta{\bf
u}_i)+v_{\mu}{\bfs\mu}_0],\Sigma_\mu^{-1}),\\
\pi({\bfs\omega}^2|{\bf{x}},{\bf{u}},{\bfs\mu},{\bfs\delta},L)&=&
\prod_{i=1}^kG(s_{\omega_i}+\frac{n+1}{2},r_{\omega_i}+\frac{1}{2}L_{i.}S_u
L_{i.}' +\frac{1}{2b_1}\delta_i^2),
\end{eqnarray*}
where $D_{u_i}=diag({\bf u}_i)$, $i=1,\cdots,n$, and
\begin{eqnarray*}
\Sigma_\delta&=&\Sigma_{i=1}^nD^2_{u_i}+{1}/{b_1}I_k,\ \ \
\Sigma_\mu=n(L'D_\omega L)+v_\mu I_k,\\
S_u&=&\Sigma_{i=1}^n({\bf x}_i-\mu-L^{-1}D_\delta{\bf u}_i)({\bf
x}_i-\mu-L^{-1}D_\delta{\bf u}_i)'.
\end{eqnarray*}
\begin{table}[t]
\centering {\caption{The value of hyperparameters under different
prior distributions.} \vspace{1mm}\label{prior}
\renewcommand{\tabcolsep}{.5pc} 
\renewcommand{\arraystretch}{.65} 
\begin{tabular}{@{}|c|  c c c| }
\hline
Hyperparameter &  Prior (\ref{indep}) & Prior (\ref{khar}) & Prior (\ref{gef0}) \\
\hline
$v_\mu$      & $1/b_2$ &  $\cdots$ &    0  \\
$s_{\omega_i}$& $b_3$   & ${\psi_i}/{2}$ &  0   \\
$r_{\omega_i}$& $b_4$   & $\frac{1}{2}L_{i.}\Psi L_{i.}'$ &  0   \\
$V_{L_i}$& $1/b_5I_k$   & $\omega_i^2\Psi $ &  {\bf 0}   \\
\hline
 \end{tabular}}
\end{table}
The full conditional for $L_{i.}^{\neq 0}$, $i=1,\cdots,k-1,$ is
\begin{eqnarray*}
\pi(L_{i.}^{\neq 0}|L\setminus L_{i.}^{\neq
0},{\bf{x}},{\bf{u}},{\bfs\mu},{\bfs\delta},{\bfs\omega}^2)=N_{||N^\prec(i)||}
(\Sigma_{(ii)}^{-1}[\omega_i^2\delta_i{M}_{i.}-{\bfs\zeta}_{(i)}],\Sigma_{(ii)}^{-1}),
\end{eqnarray*}
where ${M}_{i.}=(M_{ij})_{\{i<j,\ (i,j)\in E\}}$ with
$M=\sum_{i=1}^n{\bf u}_i{\bf x}_i'$. Additionally, $\Sigma_{(ii)}$
and ${\bfs\zeta}_{(i)}$ are submatrices of
$\Sigma^{(i)}=\omega_i^2S+V_{L_i}$ with $S=\sum_{i=1}^n({\bf
x}_i-\mu)({\bf x}_i-\mu)'$ such that
${\bfs\zeta}_{(i)}=(\Sigma^{(i)}_{ij})_{\{i<j,\ (i,j)\in E\}}$ and
$\Sigma_{(ii)}=(\Sigma^{(i)}_{kl})_{\{(k,l)|\ (k,l)\in E, \
k,l>i\}}$. Note that the resulting posterior distributions under
different values for hyperparameters can be determined based on
Table \ref{prior}.
\section{Simulation study}
The SGDG model introduces the extra parameter ${\bfs\dl}$ beyond the
parametrization of the usual Gaussian graphical model, so now we
want to examine to what extent information on this parameter can be
recovered from data. We assign a diffuse prior on ${\bfs\dl}$ using
$b_1=100$, and standard noninformative priors (\ref{gef0}) for the
other parameters. We  generate data from the SGDG model
(\ref{hirar}) with $k=3$ variables under the neighborhood graph
(\ref{lin.g}). We use a sample size of $n=200$ with ${\bfs\mu}=5{\bf
1}_3$ and $D_\om=I_3$. We focus on the skewness parameters $\dl_1$,
$\dl_2$ and $\dl_3$ and two nonzero elements $L_{12}$ and $L_{23}$
of lower triangular matrix $L$ as inference is most challenging for
these parameters. The  data sets based on different values for these
parameters has been determined as follows:
\begin{description}
    \item[Case A:] Four data sets generated by
$L_{12}=L_{23}=-0.5$ and with four different values of skewness
parameters given by $\dl_1=\dl_2=\dl_3=-1,1,2,3$
    \item[Case B:] Four data sets generated by $\dl_1=\dl_2=\dl_3=2$ and
    $L_{12}=L_{12}=-1,-0.5,0.5,1$
    \item[Case C:] One data set generated with $\dl_1=3$, $\dl_2=-2$,
$\dl_3=-4$, $L_{12}=-0.5$ and $L_{23}=0.5$.
\end{description}
The motivation for choosing the parameters in case C is that for a
given observation vector ${\bf X}=(X_1,X_2,X_3)$, we have
\begin{eqnarray*}
X_1&\stackrel{d}{=}&3|U_1|-|U_2|+|U_3|+V_1+0.5V_2-0.25V_3,\\
X_2&\stackrel{d}{=}&-2|U_2|+2|U_3|+V_2-0.5V_3,\\
X_3&\stackrel{d}{=}&-4|U_3|+V_3.
\end{eqnarray*}
where  ${\bf U}=(U_1,U_2,U_3)', {\bf
V}=(V_1,V_2,V_3)'\stackrel{iid}{\sim}N_3({\bf0},I_3)$. Hence, we can
easily see that $(X_1,X_2,X_3)'\stackrel{d}{=}(X_1,-X_2,X_3)'$, so
the marginal density for $X_2$ is symmetric.
\begin{figure}
\centerline{
\includegraphics[width=150mm,height=47mm]{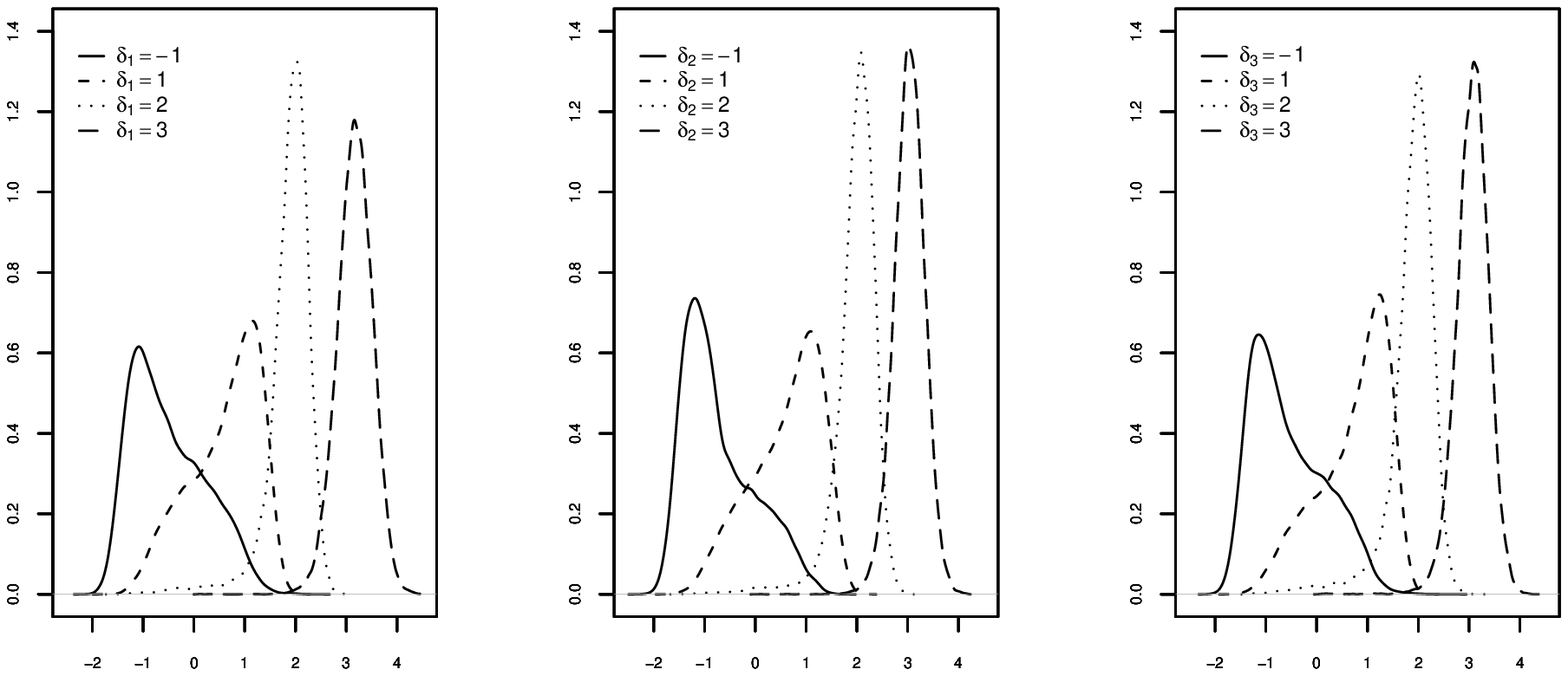}}
\caption{Posterior distributions for $\dl_1$, $\dl_2$ and $\dl_3$
under four simulated data sets in Case A. The legends indicate the
values of $\dl_i$'s used to generate data sets.}\label{alpha}
\end{figure}

\begin{figure}
\centerline{
\includegraphics[width=115mm,height=55mm]{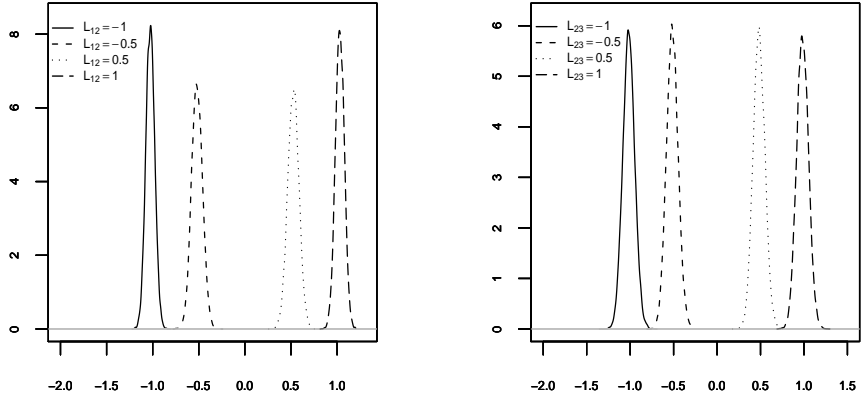}}
\caption{Posterior distributions for $L_{12}$ and $L_{23}$ under
four simulated data sets in Case B. The legends indicate the values
of $L_{12}$ and $L_{23}$ used to generate the data sets.}\label{L}
\end{figure}

\begin{figure}
\centerline{
\includegraphics[width=115mm,height=55mm]{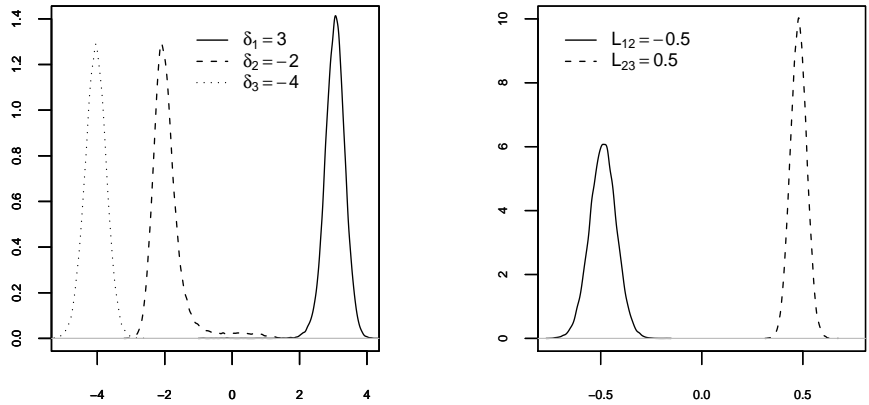}}
\caption{Posterior distributions for (left figure) $\dl_1$, $\dl_2$,
$\dl_3$ and (right figure) $L_{12}$ and $L_{23}$ under simulated
data in Case C. The legends indicate the values of these parameters
used to generate the data.}\label{alpha.L}
\end{figure}

Our results are based on a MCMC chain of length 500000 with a
burn-in  of 100000, using the block MCMC algorithm in Section 5.
Figure \ref{alpha} displays the posterior distributions for $\dl_1$,
$\dl_2$ and $\dl_3$ under four data sets introduced in Case A. These
figures clearly indicate that the data allow for meaningful
inference on ${\bfs\dl}$ since the posterior distributions assign a
large mass to neighborhoods of the values used to generate the data,
specially for larger values of the skewness.  Figure \ref{L}
displays the  posterior inference for $L_{12}$ and $L_{23}$ under
four data sets in Case B.  Figure \ref{alpha.L} displays the
posterior distributions for (left figure) $\dl_1$, $\dl_2$, $\dl_3$
and (right figure) $L_{12}$ and $L_{23}$ under a data set introduced
in Case C which indicates the data clearly allows for good
inference.
\section{Case studies}
In this section, we apply our approach to two data sets: student's
mathematics  marks from Mardia et al. (1979) and the carcass data
from Busk et al. (1999). We  compare the results with those obtained
from the Gaussian model. Our results are based on a MCMC chain of
length 700000 with a burn-in  200000.  All results in this section
were computed  under prior (\ref{indep})  with $b_1=100$,
${\bfs\mu}_0={\bf 0}$, $b_2=10^4$, $b_3=b_4=10^{-6}$ and $b_5=100$
which gives us a diffuse prior.
\subsection{Mathematics Marks}
\begin{figure}
\centerline{
\includegraphics[width=45mm,height=35mm]{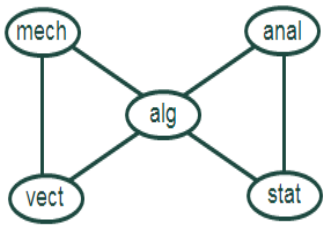}}
\caption{Neighborhood graph of mathematics marks.}\label{math.g}
\end{figure}

\begin{figure}
\centerline{
\includegraphics[width=140mm,height=100mm]{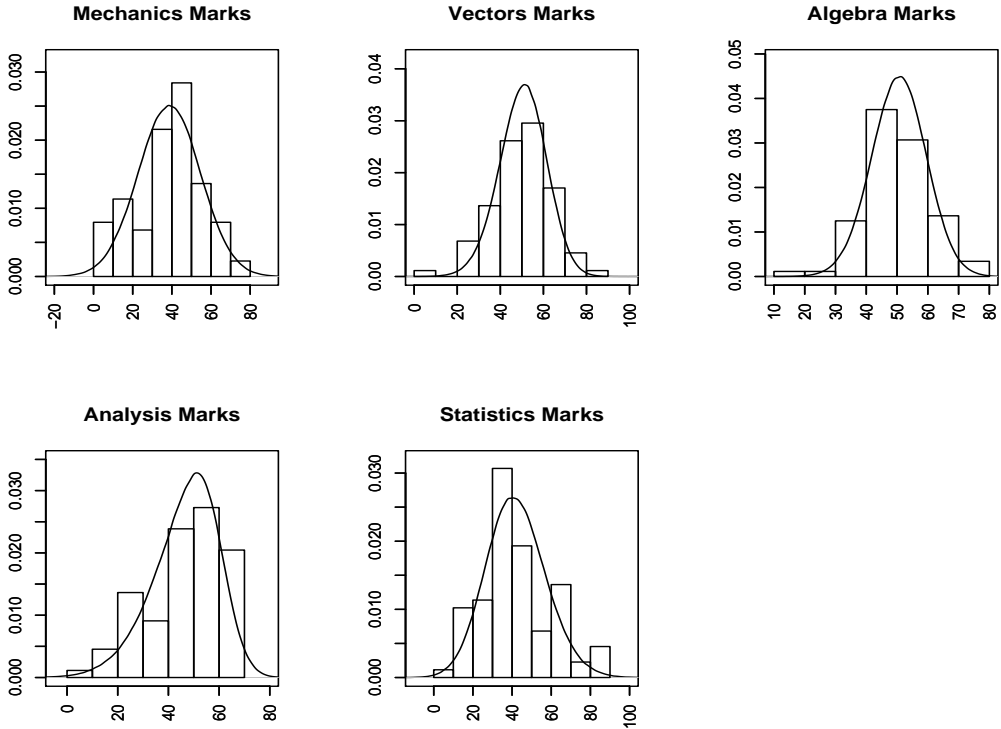}}
\caption{Histograms of mathematics marks  and fitted models to
those.}\label{hist}
\end{figure}

\begin{table}[t]
\centering {\caption{The sample partial correlation matrix of the
mathematics marks.} \vspace{1mm}\label{PC}
\renewcommand{\tabcolsep}{.5pc} 
\renewcommand{\arraystretch}{.65} 
\begin{tabular}{@{}c|  c c c  c c}
             & Mechanics& Vectors& Algebra & Analysis & Statistics\\
             \hline\\
  Mechanics  & 1&  0.33&  0.23 & 0.00  & 0.02  \\
  Vectors    &  & 1    &  0.28 & 0.08  & 0.02  \\
  Algebra    &  &      & 1     &  0.43 &  0.36 \\
  Analysis   &  &      &       & 1     &  0.25 \\
  Statistics &  &      &       &       & 1 \\
 \end{tabular}}
\end{table}
This data set come from Mardia et al. (1979), and consists of
examination marks of 88 students in the five subjects mechanics,
vectors, algebra, analysis and statistics. Mechanics and vectors
were closed book examinations and the reminder were open book. All
variables are measured on the same scale (0-100). Table \ref{PC}
displays sample partial correlation matrix between these variables.
The element in the upper righthand block are all near zero which it
means that we can consider mechanics and analysis conditionally
independent  on the other remaining variables, as are mechanics and
statistics, vectors and statistics and finally vectors and analysis.
These assumptions have been described in neighborhood graph in
Figure \ref{math.g} as suggested by Whittaker (1990). For
exploratory purpose, the histograms of variables are plotted in
Figure \ref{hist}. These histograms suggest that analysis and
statistics marks have skewed distributions in left and right,
respectively.  We used the following ordering for our analysis:
$\{$Mechanics, Vectors, Algebra, Analysis, Statistics$\}$.
The posterior mean (standard deviation) estimates of nonzero
elements of upper triangular matrix $L$ under GG and SGDG model  are
as follows
\begin{eqnarray*}
\scriptsize\hat L_{GG}&=&\left(%
\begin{array}{ccccc}
  1& -0.46_{(0.15)}& -0.55_{(0.18)}  & 0.00  & 0.00  \\
   & 1& -0.75_{(0.11)}  &  0.00 &  0.00 \\
   &  & 1 & -0.35_{(0.06)}  & -0.23_{(0.05)}  \\
   &  &   & 1 & -0.52_{(0.07)}  \\
   &  &   &   & 1 \\
\end{array}%
\right),\\ \scriptsize
\hat L_{SGDG}&=&\left(%
\begin{array}{ccccc}
  1& -0.46_{(0.14)}& -0.55_{(0.17)}  &  0.00 & 0.00  \\
   & 1& -0.76_{(0.09)}  &  0.00 &  0.00 \\
   &  & 1 & -0.32_{(0.06)}  & -0.24_{(0.05)}  \\
   &  &   & 1 & -0.44_{(0.05)}  \\
   &  &   &   & 1 \\
\end{array}%
\right).
\end{eqnarray*}
The posterior mean estimates of the other parameters  is shown in
Table \ref{est.math}. Recall that $\delta_i$ is not skewness
parameter of $i$th variable only, but the $\delta_i$'s in Table
\ref{est.math} can be approximately interpreted as skewness
parameters of the rows of $\hat L_{SGDG}X$ where $X$=(Mechanics,
Vectors, Algebra, Analysis, Statistics).  Figure \ref{hist} shows
the histograms of variables with the fitted SGDG model. The fits
seems adequate.

To compare the GG and SGDG model, we computed the Bayes factor using
the modified harmonic mean estimator $\hat p_4$ of Newton and
Raftery (1994). The Bayes factor in favor of the SGDG model is
$3\times10^{37}$, which indicates overwhelming support for the SGDG
model. We also tried  other vertex-orderings corresponding to a
perfect vertex elimination scheme, but the chosen ordering  in the
first of this section (i.e. $\{$Mechanics, Vectors, Algebra,
Analysis, Statistics$\}$) was supported by the Bayes factor.
Additionally, note that although the SGDG model depends on the
ordering of the vertices, it give a better fit in compare with the
corresponding GG model over all vertex-orderings corresponding to a
perfect vertex elimination scheme.
\begin{table}[tp]
\centering { \caption{Mathematics marks: Posterior means (standard
deviation)  for parameters.} \vspace{1mm}\label{est.math}
\renewcommand{\tabcolsep}{.5pc} 
\renewcommand{\arraystretch}{.65} 
 \begin{tabular}{@{}c  c  c c| c  c}
 \hline\\
&&  SGDG  &&~~GG &\\
 i  &\cline{1-5}
 & ${\bfs\beta}$& ${\bfs\om}^2$ & ${\bfs\dl}$~~~~~& ${\bfs\beta}$& ${\bfs\om}^2$ \\
  \hline\\
  1   & $45.83_{(5.75)}$ & $0.0057_{(0.0015)}$ & $0.38_{(6.15)}$   & $39.16_{(1.83)}$  & $0.0051_{(0.0008)}$  \\
  2   & $61.21_{(4.83)}$ & $0.0235_{(0.0174)}$ & $-9.17_{(5.63)}$  & $50.68_{(1.44)}$  & $0.0091_{(0.0014)}$  \\
  3   & $55.18_{(3.27)}$ & $0.0469_{(0.0279)}$ & $-5.51_{(4.15)}$  & $50.64_{(1.12)}$  & $0.0211_{(0.0032)}$  \\
  4   & $56.17_{(2.30)}$ & $0.2131_{(0.3696)}$ & $-18.29_{(1.88)}$ & $46.77_{(1.56)}$  & $0.0071_{(0.0011)}$  \\
  5   & $30.53_{(3.68)}$ & $0.0065_{(0.0025)}$ & $14.54_{(3.89)}$  & $42.25_{(1.88)}$  & $0.0034_{(0.0005)}$  \\
 \hline
 \end{tabular}}
\end{table}
\subsection{Carcass Data}
\begin{figure}
\centerline{
\includegraphics[width=55mm,height=55mm]{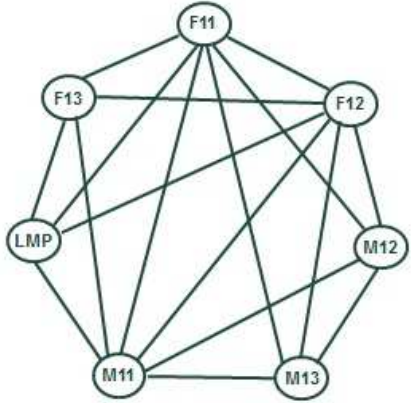}}
\caption{Neighborhood graph of carcass data.}\label{car.G}
\end{figure}

The carcass data from gRbase package of R contains measurements of
the thickness of meat and fat layers together with the lean meat
percentage of 344 slaughter carcasses at three Danish slaughter
houses. Seven variables has been defined in this data set as
follows:
\begin{itemize}
    \item F11, F12, F13: Thickness of fat layer at 3 different locations on the
          back of the carcass.
    \item M11, M12, M13: Thickness of meat layer at 3 different locations on the
          back of the carcass.
    \item LMP: Lean meat percentage determined by dissection
\end{itemize}
This data set has been used for estimating the parameters in a
prediction formula for prediction of lean meat percentage on the
basis of the thickness measurements on the carcass. Data are
described in detail in Busk et al. (1999). Hojsgaard et al. (2012)
provided some neighborhood graphs for these variables in term of
different model selection methods. Based on the BIC criterion, they
proposed the neighborhood graph in Figure \ref{car.G}.
\begin{table}[tp]
\centering { \caption{Carcass Data: Posterior means (standard
deviation)  for parameters.} \vspace{1mm}\label{est.car}
\renewcommand{\tabcolsep}{.5pc} 
\renewcommand{\arraystretch}{.65} 
 \begin{tabular}{@{}c  c  c c| c  c}
 \hline\\
&&  SGDG  &&~~GG &\\
 i  &\cline{1-5}
 & ${\bfs\beta}$& ${\bfs\om}^2$ & ${\bfs\dl}$~~~~~& ${\bfs\beta}$& ${\bfs\om}^2$ \\
  \hline\\
  1   & $51.67_{(2.62)}$ & $0.152_{(1.41)}$ & $0.54_{(0.025)}$ & $52.01_{(0.41)}$ & $0.133_{(0.010)}$ \\
  2   & $55.89_{(2.38)}$ & $0.112_{(1.70)}$ & $0.87_{(0.021)}$ & $55.70_{(0.35)}$ & $0.096_{(0.007)}$ \\
  3   & $52.44_{(2.15)}$ & $0.047_{(2.46)}$ & $1.32_{(0.008)}$ & $51.98_{(0.31)}$ & $0.041_{(0.003)}$ \\
  4   & $63.81_{(0.96)}$ & $0.326_{(0.91)}$ & $-2.06_{(0.085)}$& $59.38_{(0.19)}$ & $0.196_{(0.015)}$ \\
  5   & $10.24_{(0.83)}$ & $0.615_{(0.79)}$ & $0.43_{(0.131)}$ & $12.95_{(0.15)}$ & $0.510_{(0.038)}$ \\
  6   & $13.46_{(0.98)}$ & $0.342_{(1.06)}$ & $0.25_{(0.062)}$ & $16.49_{(0.18)}$ & $0.293_{(0.023)}$ \\
  7   & $10.96_{(0.54)}$ & $0.286_{(0.66)}$ & $3.75_{(0.090)}$ & $13.96_{(0.16)}$ & $0.112_{(0.009)}$ \\
 \hline
 \end{tabular}}
\end{table}
\begin{figure}
\centerline{
\includegraphics[width=145mm,height=190mm]{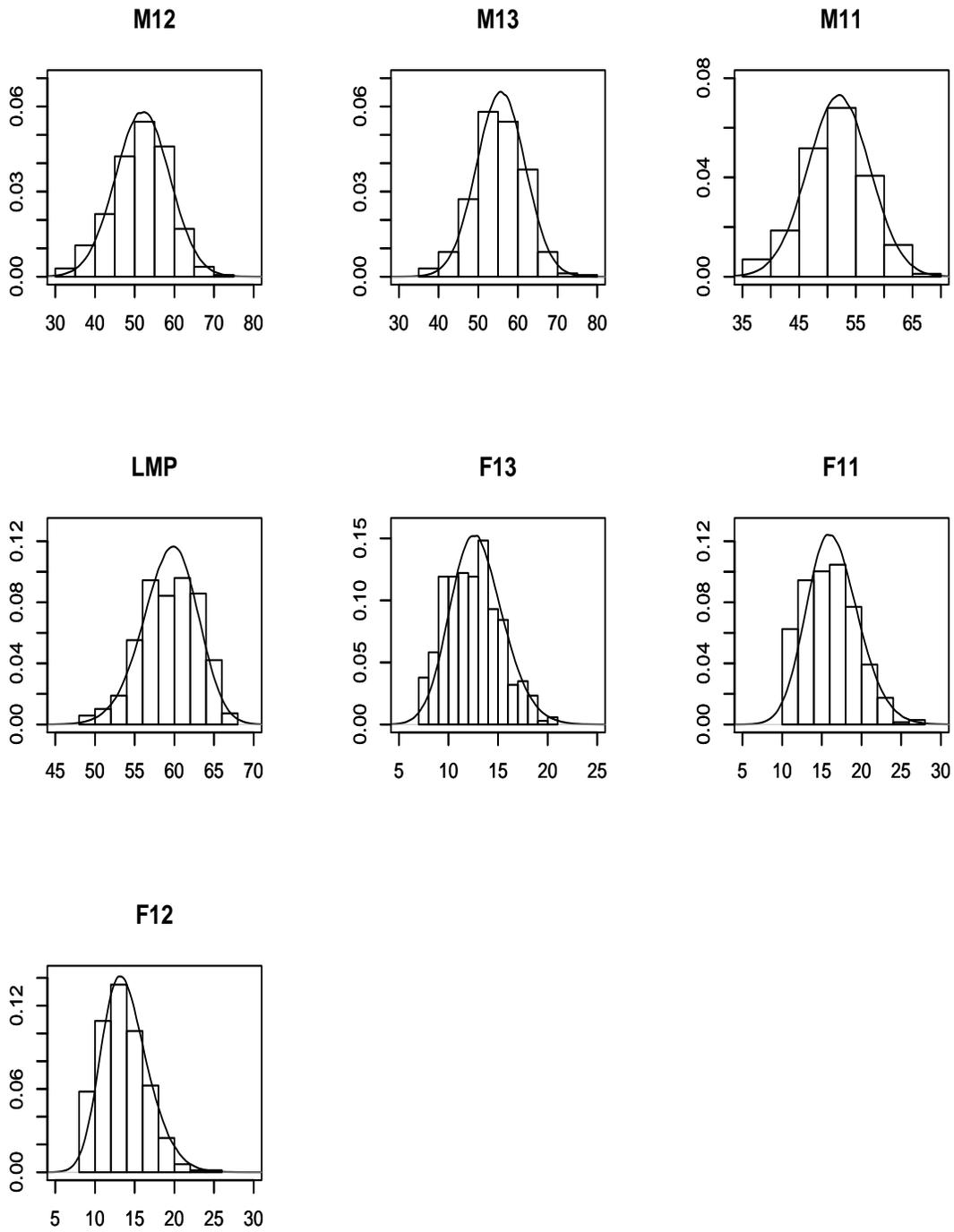}}
\caption{Histograms of carcass data  and fitted models to
those.}\label{hist.car}
\end{figure}

We used first the ordering: $\{M12, M13, M11, LMP,$ $F13, F11,
F12\}$. A summary of the posterior inference on the parameters in
the models is provided in Table \ref{est.car}. We also present two
quantities that are  directly comparable between the SGDG and GG
model. In Figure \ref{hist.car}, we have displayed the fitted SGDG
model to this data set. The SGDG model seems to fit well and
includes skewness for LMP, F13, F11 and F12. The Bayes factor in
favor of the SGDG versus the GG model was estimated to
$1.7\times10^{8}$. We also tried  an alternative orderings for our
analysis, but they gave all a worse fit.

\section{Conclusions}
In this paper, we have proposed a novel decomposable graphical model
to accommodate skew-Gaussian decomposable graphical model (SGDG).
The SGDG model reflect conditional independencies among the
components of the multivariate closed skew normal random vector with
respect to a decomposable graph, and includes the Gaussian graphical
models as a particular case.

We develop a family of conjugate prior distributions for precision
matrix, and derive a condition which ensures propriety of the
posterior distribution corresponding to the standard noninformative
priors for mean vector, precision parameters and lower triangular
matrix  as well as proper prior for
skewness parameter.
We successfully apply our new model to two data sets, showing great
improvement on the Gaussian graphical model. 

Since the SGDG model satisfy the conditional independence property
only under a decomposable graph, so an extension of the SGDG model
for other graphs requires further research.
\newpage
\ \ \ \ \ \ \ \ \ \ \ \ \ \ \ \ \ \ \ \ \ \ \ \ \ \ \ \ \ \ \ \ \ \
{\bf APPENDIX: PROOFS} \\
\\
To prove  Theorem 1, we use the following
factorization criterion.\\
\\
{\bf Lemma 4:} Two random variables $X$ and $Y$ are called
conditionally independent given $Z$ iff there exist some functions
$f$ and $g$ such that $\pi(x,y,z)=f(x,z)g(y,z)$ (see Rue and Held
(2005)).\\
\\
{\bf Proof of Theorem 1:} We partition ${\bf X}$ as $(X_i,X_j,{\bf
X}_{-ij})$ and then use the multivariate version of the
factorization criterion on $\pi(x_i,x_j,{\bf x}_{-ij})$. Fix $i\neq
j$ and assume $\bfs\mu={\bf0}$ without loss of generality. From
(\ref{sgmrf.decom}) we get
\begin{eqnarray*}
\pi(x_i,x_j,{\bf x}_{-ij})\propto\exp(-\frac{1}{2}\sum_{m,l}x_m\
Q_{ml}\ x_l)\prod_{r=1}^k\ \Phi_1(\al_r\kappa_r L_{r.}'{\bf x};\ 0,\
1),
\end{eqnarray*}
where $\kappa_r$ is root square of the $r$'th diagonal element of
$D_\kappa$ and $L_{r.}$ is $r$'th row of $L$. Let $i<j$ without loss
of generality. Now, we define  set
\begin{eqnarray*}
I&=&\{r:\ \  L_{ri}\neq0\ \ and\ \ L_{rj}=0\  \ for\ \ r\leq i\}
\end{eqnarray*}
Then, due to the  ordering of the vertices corresponds to a perfect
vertex elimination scheme and with respect to Definition 2, we know
if $\upsilon<i<j$ then the neighbors of vertex $\upsilon$ can not be
vertex $i$ and $j$ together if and only if vertices $i$ and $j$ are
not neighbors. Hence, we have
\begin{eqnarray*}
\pi(x_i,x_j,{\bf
x}_{-ij})&\propto&\exp(-\frac{1}{2}x_ix_j(Q_{ij}+Q_{ji})-\frac{1}{2}\sum_{\{m,l\}\neq
\{i,j\}}x_m\ Q_{ml}\ x_l)\\
&\times&\prod_{r\in I}\ \Phi_1(\al_r\kappa_r L_{r.}'{\bf x};\ 0,\
1)\prod_{r\notin I}\ \Phi_1(\al_r\kappa_r L_{r.}'{\bf x};\ 0,\
1).\label{sepret}
\end{eqnarray*}
Second term  does not involve $x_ix_j$ while first term  involves
$x_ix_j$ iff $Q_{ij}\neq 0$. Thus,  one example for functions $f$
and $g$ in lemma can be defined as
\begin{eqnarray*}
f(x_i,{\bf x}_{-ij})&=&\exp(-\frac{1}{2}\sum_{\{m,l\}\neq j}x_m\
Q_{ml}\ x_l) \prod_{r\in I}\ \Phi_1(\al_r\kappa_r
L_{r.}'{\bf x};\ 0,\ 1)\\
g(x_j,{\bf x}_{-ij})&=&\exp(-\frac{1}{2}\sum_{\{m,l\}\neq i}x_m\
Q_{ml}\ x_l)\prod_{r\notin I}\ \Phi_1(\al_r\kappa_r L_{r.}'{\bf x};\
0,\ 1).
\end{eqnarray*}
iff $Q_{ij}\neq 0$. The claim then follows. $\blacksquare$.\\
\\
{\bf Proof of Theorem 2:}  At first, we recall if $G$ be a
nondecomposable undirected graph, then it is possible that the zero
elements of precision matrix $Q$ are not reflected in its modified
cholesky decomposition $L$. To prove this theorem, we proceed
similar to Theorem 1. Fix $i<j$ and assume $\bfs\mu={\bf0}$ without
loss of generality. Here, we know when  $F(i,j)$ separates vertices
$i$ and $j$, then $L_{ij}=0$ and it is not possible to have a pass
from $i$ to $j$ with passing from some vertex  that the number of
all of them are smaller than $i$. Thus, if $\upsilon<i<j$, then
$\upsilon$ can not be the neighbors of vertex $i$ and $j$ together.
Hence, we can write $\pi(x_i,x_j,{\bf x}_{-ij})$ similar to
(\ref{sepret}) with same definition for set $I$. The rest of the
proof is similar to that of Theorem 1.\ \ $\blacksquare$\\
\\
{\bf Proof of Theorem 3:} At first, we denote $L_{i\cdot}$  as
$i$'th row  of $L$. Now, we have
\begin{eqnarray*}
&&\int\prod_{i=1}^k
(\omega^2_i)^{\psi_i/2-1}\exp\{-\frac{1}{2}tr((L'D_\omega
L)\Psi)\} dD dL\\
&=&\int\prod_{i=1}^k(\omega^2_i)^{\psi_i/2-1}\exp\{-\frac{1}{2}(L\Psi
L')_{ii}\omega^2_i\} dD dL
\propto\int\prod_{i=1}^k\frac{1}{(L\Psi L')_{ii}^{\psi_i/2}}dL\\
&\propto&\prod_{i=1}^{k-1}\int\frac{1}{(L_{i\cdot}\Psi
L_{i\cdot}')^{\psi_i/2}}dL_{i\cdot}=\prod_{i=1}^{k-1}\int_{\Re^{||N^\prec(i)||}}\frac{1}
{(L_{i\cdot}^*\Psi^{(i)}
L_{i\cdot}^{*'})^{\psi_i/2}}dL_{i\cdot}^{\neq 0}
\end{eqnarray*}
where $L_{i\cdot}^*=(1\ \ L_{i\cdot}^{\neq 0})'$. Also, $\Psi^{(i)}$
is a submatrix of $\Psi$ corresponding to the elements of
$L_{i\cdot}^*$. We are now in the same line with Khare and
Rajaratnam (2011) and similarly, we can show that the integral is
finite if $\psi_i>||N^\prec(i)||$.\ \
$\blacksquare$\\
\\
{\bf Proof of Theorem 4}: Under the introduced reparemeterizations
in hierarchical model (\ref{hirar}), the conditional density of
${\bf X}$ becomes
\begin{eqnarray*}
p({\bf
x}|{\bfs\mu},{\bfs\omega}^2,{\bfs\delta},L)=(\frac{2}{\pi})^{k/2}|D_g|^{\frac{1}{2}}
e^{-\frac{1}{2}({\bf x}-{\bfs\mu})'Q_g({\bf x}-{\bfs\mu})}
\prod_{i=1}^k\Phi(\dl_i\omega_i\sqrt {g(\omega_i,\delta_i)} L({\bf
x}-{\bfs\mu});0,1),
\end{eqnarray*}
where $g(\omega_i,\delta_i)=\frac{\o2_i}{1+\dl_i^2\o2_i}$,
$D_g=diag(g(\omega_1,\delta_1)),\cdots,g(\omega_k,\delta_k)$ and
$Q_g=L'D_gL$. Now, since $0\leq\Phi(\cdot)\leq 1$, we have
\begin{eqnarray*}
p({\bf x})&=&\int_{\Re^N}\int_{\Re^k}\int_{\Re_+^k}\int_{\Re^k}
p({\bf x}|{\bfs\mu},{\bfs\omega}^2,{\bfs\delta},L)\pi({\bfs\mu})
\pi({\bfs\delta}|{\bfs\omega}^2)\pi(L,D_\omega)d{\bfs\mu}
d{\bfs\omega}^2 d{\bfs\delta}dL\\
&<&\int_{\Re^N}\int_{\Re^k}\int_{\Re_+^k}\int_{\Re^k}|D_g|^{\frac{n}{2}}
e^{-\frac{1}{2}\sum_{i=1}^n({\bf x}_i-{\bfs\mu})'Q_g({\bf
x}-{\bfs\mu}_i)}
\pi({\bfs\delta}|{\bfs\omega}^2)\pi(L,D_\omega)d{\bfs\mu}
d{\bfs\omega}^2 d{\bfs\delta}dL\\
&=&K_1\int_{\Re^k} f_N^k(\bfs\mu|\bar{\bf x},(nQ_g)^{-1})d{\bfs\mu}\\
&\times&\int_{\Re^N}\int_{\Re^k}\int_{\Re_+^k}|D_g|^{\frac{n}{2}}|Q_g|^{-\frac{1}{2}}
e^{-\frac{1}{2}\sum_{i=1}^n({\bf x}_i-\bar{\bf x})'Q_g({\bf
x}_i-\bar{\bf x})} \pi({\bfs\delta}|{\bfs\omega}^2)\pi(L,D_\omega)
d{\bfs\omega}^2 d{\bfs\delta}dL
\end{eqnarray*}
where  $N=\sum_{i=1}^{k-1}||N^\prec(i)||$, $\bar{\bf
x}=\frac{1}{n}\sum_{i=1}^n{\bf x}_i$ and $K_1$ is a constant. If we
define $S=\sum_{i=1}^n({\bf x}_i-\bar{\bf x})({\bf x}_i-\bar{\bf
x})'$ and using variable transformation ${\lam}_i=\dl_i{\omega}_i$,
$i=1,\cdots,k$,   under $n\geq 2$, we get
\begin{eqnarray*}
p({\bf x})&<&K_2\int_{\Re^N}\int_{\Re^k}\int_{\Re_+^k}
\prod_{i=1}^k(\frac{1}{1+\dl_i^2\o2_i})^{\frac{n-1}{2}}\prod_{i=1}^k(\o2_i)^{\frac{n}{2}-1}
e^{-\frac{1}{2}tr(Q_gS)}e^{-\frac{1}{2b_2}\sum_{i=1}^k\o2_i\dl_i^2}
d{\bfs\omega}^2 d{\bfs\delta}dL\\
&=&K_2\int_{\Re^N}\int_{\Re^k}\int_{\Re_+^k}
\prod_{i=1}^k\{(\frac{1}{1+\lam_i^2})^{\frac{n-1}{2}}e^{-\frac{1}{2b_2}\lam_i^2}
(\o2_i)^{\frac{n-1}{2}-1} e^{-\frac{1}{2}\frac{(LS
L')_{ii}}{1+\lam_i^2}\omega^2_i}\} d{\bfs\omega}^2 d{\bfs\lam}dL\\
&\propto&K_2\int_{\Re^N}\prod_{i=1}^k\frac{1}{(LS
L')_{ii}^{\frac{n-1}{2}}}dL=\prod_{i=1}^{k-1}\int_{\Re^{||N^\prec(i)||}}\frac{1}
{(L_{i\cdot}^*\Psi^{(i)}
L_{i\cdot}^{*'})^{\frac{n-1}{2}}}dL_{i\cdot}^{\neq 0}
\end{eqnarray*}
where $K_2$ is  another constant. Similarly to the proof of Theorem
3, we
can show that the integral is finite if $n\geq \max\{||N^\prec(i)||\}+2$.\ \ $\blacksquare$\\


\end{document}